\documentclass[10pt, twoside, openright]{scrartcl}

\usepackage[ngerman, english]{babel}
\usepackage[a4paper,%
            top=2.5cm,bottom=3.5cm,right=2.5cm,left=2.5cm,bindingoffset=5mm]{geometry}

\usepackage[utf8]{inputenc}
\usepackage[T1]{fontenc}

\usepackage{amsmath}    
\usepackage{amssymb}

\usepackage{dsfont}     
\usepackage{amsfonts}

\usepackage{graphicx}   
\usepackage{subfigure}  

\usepackage[sc]{mathpazo}   
\usepackage{booktabs}   
\usepackage{dcolumn}    
\usepackage{bm}         

\usepackage{colortbl}   

\usepackage[authoryear,longnamesfirst]{natbib}     
\bibpunct{(}{)}{,}{a}{}{,}

\usepackage{setspace}
\usepackage{rotating}

\usepackage{ulem}      
\usepackage{fancyhdr}
\usepackage{fancyvrb}  

\usepackage{%
  ellipsis, 
  ragged2e, 
 marginnote,
}

\usepackage{keyval}

\usepackage{wallpaper}   

\usepackage{float}

\usepackage{framed}
\definecolor{shadecolor}{cmyk}{0.02,0.02,0.02,0.02}

\usepackage[toc,page]{appendix}


\setcapindent{2em}
\addtokomafont{caption}{\small}
\setkomafont{captionlabel}{\bfseries}

\usepackage{xr}
\newcommand{\citetappendix}[1]{Online Resource 1 (Section~\ref{#1})}
\newcommand{\tabappendix}[1]{(Online Resource 1; Table~\ref{#1})}
\newcommand{\citepappendix}[1]{(Online Resource 1; Section~\ref{#1})}
\RequirePackage[colorlinks,citecolor=blue,urlcolor=blue]{hyperref}
\usepackage{subfigure}
\newcommand{\sidecaption}{\relax}
\newenvironment{acknowledgements}{\par\subsection*{Acknowledgements}}{\par}{}

\newcommand{\mbf}[1]{\boldsymbol{#1}}

\newcommand{\COMMENT}[1]{}

\newcommand{\R}[1]{\texttt{#1}}

\DeclareMathOperator{\diag}{diag}

\makeatletter
\newcommand\floatc@simplerule[2]{{\@fs@cfont #1} #2\par}
\newcommand\fs@simplerule{\def\@fs@cfont{\bfseries}\let\@fs@capt\floatc@simplerule
  \def\@fs@pre{\hrule height1.2pt depth0pt \kern4pt}%
  \def\@fs@mid{\vspace*{0.5em} \hrule height.3pt depth0pt \vspace*{0.8em} \kern4pt}%
  \def\@fs@post{\kern4pt \hrule height1.2pt depth0pt \kern4pt \relax}%
  \let\@fs@iftopcapt\iftrue}
\makeatother

\renewcommand{\theenumi}{(\alph{enumi})}

\renewcommand{\theenumii}{(\alph{enumi}\arabic{enumii})}

\newcommand{\roedeer}[0]{(\textit{Capreolus capreolus})}

\newcommand{\mbmono}[0]{\texttt{`mboost'}}
\newcommand{\mgcv}[0]{\texttt{`mgcv'}}
\newcommand{\scam}[0]{\texttt{`scam'}}

\newcommand{\uv}{\mathbf{u}}

\newcommand{\vv}{\mathbf{v}}

\newcommand{\V}{\mathbf{V}}

\newcommand{\x}{\mathbf{x}}

\newcommand{\I}{\mathbf{I}}

\newcommand{\B}{\mathbf{B}}
\newcommand{\K}{\mathbf{K}}

\newcommand{\D}{\mathbf{D}}

\newcommand{\bv}{\mbf{\beta}}

\newcommand{\Prob}{\mathbb{P}}

\newcommand{\Eqref}[1]{(Eq.~\ref{#1})}

\usepackage[all]{hypcap} 

\begin{document}\sloppy

\title{A Unified Framework of Constrained Regression}

\author{\large Benjamin Hofner\thanks{E-mail: {benjamin.hofner@fau.de}}
  \footnote{Institut f\"ur Medizininformatik, Biometrie und Epidemiologie,
    Friedrich-Alexander-Universit\"at Erlangen-N\"urnberg, Waldstra{\ss}e 6,
    D-91054 Erlangen, Germany} \and \large Thomas Kneib\footnote{Lehrstuhl f\"ur Statistik,
    Georg-August-Universit\"at G\"ottingen, Germany} \and \large Torsten Hothorn
  \footnote{Institut f\"ur Sozial- und Pr\"aventivmedizin, Abteilung
    Biostatistik, Universit\"at Z\"urich, Switzerland}\\[1em]
}

\date{\normalsize \textbf{Technical Report: \today}}

\maketitle

\begin{abstract}
Generalized additive models (GAMs) play an important role in modeling and
  understanding complex relationships in modern applied statistics. They allow
  for flexible, data-driven estimation of covariate effects. Yet researchers
  often have a priori knowledge of certain effects, which might be monotonic or
  periodic (cyclic) or should fulfill boundary conditions. We propose a unified
  framework to incorporate these constraints for both univariate and bivariate
  effect estimates and for varying coefficients. As the framework is based on
  component-wise boosting methods, variables can be selected intrinsically, and
  effects can be estimated for a wide range of different distributional
  assumptions. Bootstrap confidence intervals for the effect estimates are
  derived to assess the models. We present three case studies from environmental
  sciences to illustrate the proposed seamless modeling framework. All discussed
  constrained effect estimates are implemented in the comprehensive \textsf{R}
  package \textbf{mboost} for model-based boosting.
\end{abstract}

\section{Introduction}
\label{sec:introduction}

When statistical models are used, certain assumptions are made, either for
convenience or to incorporate the researchers' assumptions on the shape of
effects, e.g., because of prior knowledge. A common, yet very strong assumption in
regression models is the linearity assumption. The effect estimate is
constrained to follow a straight line. Despite the widespread use of linear
models, it often may be more appropriate to relax the linearity assumption.

Let us consider a set of observations $(y_i, \mbf{x}^\top_i ), i = 1, \ldots,
n,$ where $y_i$ is the response variable and $\mbf{x}_i = (x_i^{(1)}, \ldots,
x_i^{(L)})^\top$ consists of $L$ possible predictors of different nature, such
as categorical or continuous covariates. To model the dependency of the response
on the predictor variables, we consider models with structured additive
predictor $\eta(\mbf{x})$ of the form
\begin{equation}
  \label{eq:structured_predictor}
  \eta(\mbf{x}) = \beta_0 + \sum_{l = 1}^{L} f_l(\mbf{x}),
\end{equation}
where the functions $f_l(\cdot)$ depend on one or more predictors contained in
$\mbf{x}$. Examples include linear effects, categorical effects, and smooth
effects. More complex models with functions that depend on multiple variables
such as random effects, varying coefficients, and bivariate effects, can be
expressed in this framework as well \citep[for details, see][]{Fahrmeir2004}.

Structured additive predictors can be used in different types of regression
models. For example, replacing the linear predictor of a generalized linear
model with \eqref{eq:structured_predictor} yields a structured additive
regression (STAR) model, where $\mathds{E}(y|\mbf{x}) = h(\eta(\mbf{x}))$ with
(known) response function $h$. However, structured additive predictors can be
used much more generally, in any class of regression models (as we will
demonstrate in one of our applications in a conditional transformation model
that allows to describe general distributional features and not only the mean in
terms of covariates).

A convenient way to fit models with structured additive predictors is given by
com\-ponent-wise functional gradient descent boosting \citep{buehlmann03}, which
minimizes an empirical risk function with the aim to optimize prediction
accuracy. In case of structured additive regression, the risk will usually
correspond to the negative log-likelihood but more general types of risks can be
defined for example for quantile regression models \citep{Fenske:2011}, or in
the context of conditional transformation models
\citep{Hothorn_Kneib_Buehlmann_2014}. The boosting algorithm is especially
attractive due to its intrinsic variable selection properties
\citep{Kneib:Hothorn:Tutz:modelchoice:2009,Hofner:unbiased:2011} and the ease of
combining a wide range of modeling alternatives in a single model specification.
Furthermore, a single estimation framework can be used for a very wide range of
distributional assumptions or even in distribution free approaches. Thus,
boosting models are not restricted to exponential family distributions.

Models with structured additive predictors offer great flexibility but typically
result in smooth yet otherwise unconstrained effect estimates $\hat{f}_l$. To
overcome this, we propose a framework to fit models with constrained structured
additive predictors based on boosting methods. We derive cyclic effects in the
boosting context and improved fitting methods for monotonic P-splines.
Bootstrap-confidence intervals are proposed to assess the fitted models.

\subsection[Application of Constrained Models]{Application of Constrained Models}
\label{sec:appl-constr-models}

In the first case study presented, we model the effect of air pollution on daily
mortality in S\~{a}o Paulo (Section \ref{sec:sao-paulo}). Additionally, we
control for environmental conditions (temperature, humidity) and model both the
seasonal pattern and the long-term trend. Furthermore, we consider the effect of
the pollutant of interest, SO$_2$. In modeling the seasonal pattern of mortality
related to air pollution, the effect should be continuous over time, and huge
jumps for effects only one day apart would be unrealistic. Thus, the first and
last days of the year should be continuously joined. Hence, we use smooth
functions with a cyclic constraint (Section~\ref{sec:cyclic-p-splines}). This
has two effects. First, it allows us to fit a plausible model as we avoid jumps
at the boundaries. Second, the estimation at the boundaries is stabilized as we
exploit the cyclic nature of the data.

From a biological point of view, it seems reasonable to expect an increase in
mortality with increasing concentration of the pollutant SO$_2$. Linear effects
are monotonic but do not offer enough flexibility in this case. Smooth effects,
on the other hand, offer more flexibility, but monotonicity might be violated.
To bridge this gap, smooth monotonic effects can be used
(Sec.~\ref{sec:monotonic-effects}). Additionally to the proposed boosting
framework, we will use the framework for constrained structured additive models
of \citet{PyaWood:scam:2014} (which is similar to ours but is restricted to
exponential family distributions) for comparison in the first case study.

In a second case study, we aim at modeling the activity of roe deer in Bavaria,
Germany, given environmental conditions, such as temperature, precipitation, and
depth of snow; animal-specific variables, such as age and sex; and a temporal
component. The latter reflects the animals' day/night rhythm as well as seasonal
patterns. We model the temporal effect as a smooth bivariate effect as the days
change throughout the year, i.e., the solar altitude changes in the course of a
day and with the seasons. Cyclic constraints for both variables (time of the day
and calendar day) should be used. Hence, we have a bivariate periodic effect
$f(t_{\text{hours}}, t_{\text{days}})$ (Section~\ref{sec:bivar-cyclic}). As male
and female animals differ strongly in their temporal activity profiles, we
additionally use sex as a binary effect modifier $f(t_{\text{hours}},
t_{\text{days}}) I_{(\text{sex = male})}$, i.e., we have a varying coefficient
surface with a cyclicity constraint for the smooth bivariate effect.
Additionally, the effects of environmental variables are allowed to smoothly
vary over time but are otherwise unconstrained.

In a third case study, we go beyond a model for the mean activity by modeling
the conditional distribution of a surrogate of roe deer activity: the number of
deer--vehicle collisions per day. In the framework of conditional transformation
models \citep{Hothorn_Kneib_Buehlmann_2014}, we fit daily distributions of the
number of such collisions and penalize differences in these distributions
between subsequent days. A monotonic constraint is needed to fit the conditional
distribution, while a cyclic constraint should be used for the seasonal effect
of deer--vehicle collisions. These two conditions yield a tensor product of two
univariate constrained effects.

\subsection[Overview of the Paper]{Overview of the Paper}

Model estimation based on boosting is briefly introduced in
Section~\ref{sec:boosting}. Monotonic effects, cyclic P-splines, and P-splines
with boundary constraints are introduced in
Section~\ref{sec:constrained-regression}, where we also introduce varying
coefficients. An extension of monotonicity and cyclicity constraints to
bivariate P-splines is given in Section~\ref{sec:constr-effects-bivar}. In
Section~\ref{sec:CIs} we sketch inferential procedures and derive bootstrap
confidence intervals for the constrained boosting framework. Computational
details can be found in Section~\ref{sec:software}. We present the three case
studies described above in Section~\ref{sec:case-studies}. An overview of past
and present developments of constrained regression models is given in
\citetappendix{sec:overview}. The definition of the Kronecker product and
element-wise matrix product is given in \citetappendix{sec:matrix-algebra}.
Details on the S\~{a}o Paulo air pollution data set and the model specification
is given in \citetappendix{sec:sao_paulo_data}, while
\citetappendix{sec:roedeer_data} gives details on the Bavarian roe deer data and
the specification used to model the activity of roe deer.
\citetappendix{sec:simulation_study} gives an empirical evaluation of the
proposed methods. \textsf{R} code to reproduce the fitted models from our case
studies is given as electronic supplement.

\section{Model Estimation Based on Boosting}\label{sec:boosting}

To fit a model with structured additive predictor
\eqref{eq:structured_predictor} by component-wise boosting \citep{buehlmann03},
one starts with a constant model, e.g., $\hat{\eta}(\x) \equiv 0$, and computes
the negative gradient $\uv = (u_1, \dots, u_n)^\top$ of the loss function
evaluated at each observation. An appropriate loss function is guided by the
fitting problem. For Gaussian regression models, one may use the quadratic loss
function, and for generalized linear models, the negative log-likelihood. In the
Gaussian regression case, the negative gradient $\uv$ equals the standard
residuals; in other cases, $\uv$ can be regarded as ``working residuals''. In
the next step, each model component $f_l, \,l = 1, \dots, L,$ of the structured
additive model~\eqref{eq:structured_predictor} is fitted separately to the
negative gradient $\uv$ by penalized least-squares. Only the model component
that best describes the negative gradient is updated by adding a small
proportion of its fit (e.g., $10\%$) to the \emph{current} model fit. New
residuals are computed, and the whole procedure is iterated until a fixed number
of iterations is reached. The final model $\hat{\eta}(\x)$ is defined as the sum
of all models fitted in this process. As only \textit{one} modeling component is
updated in each boosting iteration, variables are selected by stopping the
boosting procedure after an appropriate number of iterations. This is usually
done using cross-validation techniques.

For each of the model components, a corresponding regression model that is
applied to fit the residuals has to be specified, the so-called base-learner.
Hence, the base-learners resemble the model components $f_l$ and determine which
functional form each of the components can take. In the following sections, we
introduce base-learners for smooth effect estimates and derive special
base-learners for fitting constrained effect estimates. These can then be
directly used within the generic model-based boosting framework without the need
to alter the general algorithm. For details on functional gradient descent
boosting and specification of base-learners, see \citet{buehl:hoth:2007} and
\citet{Hofner:mboost:2014}.

\section{Constrained Regression}
\label{sec:constrained-regression}

\subsection{Estimating Smooth Effects}
\label{sec:p-splines}

For the sake of simplicity in the remainder of this paper, we will consider an
arbitrary continuous predictor $x$ and a single base-learner $f_l$ only when we
drop the function index $l$. To model smooth effects of continuous variables, we
utilized penalized B-splines (i.e., P-splines). These were introduced by
\citet{eilers1996} for nonparametric regression and were later transferred to
the boosting framework by \citet{Schmid:Hothorn:boosting-p-Splines}. Considering
observations $\x = (x_1, \ldots, x_n)^\top$ of a single variable $x$, a
non-linear function $f(x)$ can be approximated as
\begin{equation*}
  f(x) = \sum_{j = 1}^J \beta_j B_j(x; \delta) = \B(x)^\top \bv,
\end{equation*}
where $B_j(\cdot; \delta)$ is the $j$th B-spline basis function of degree
$\delta$. The basis functions are defined on a grid of $J - (\delta - 1)$ inner
knots $\xi_1, \ldots, \xi_{J - (\delta - 1)}$ with additional boundary knots
(and usually a knot expansion in the boundary knots) and are combined in the
vector $\B(x) = (B_1(x), \dots, B_J(x))^\top$, where for simplicity $\delta$ was
dropped. For more details on the construction of B-splines, we refer the reader
to \citet{eilers1996}. The function estimates can be written in matrix notation
as $\hat{f}(\x) = \B \hat{\bv}$, where the design matrix $\B = (\B(x_1), \ldots,
\B(x_n))^\top$ comprises the B-spline basis vectors $\B(x)$ evaluated for each
observation $x_i$, $i = 1, \ldots, n$. The function estimate $\hat{f}(\x)$ might
adapt the data too closely and might become too erratic. To enforce smoothness
of the function estimate, an additional penalty is used that penalizes large
differences of the coefficients of adjacent knots. Hence, for a continuous
response $\uv$ (here the negative gradient vector), we can estimate the function
by minimizing a penalized least-squares criterion
\begin{equation}\label{eq:pen_LS_pspline}
  \mathcal{Q}(\bv) = (\uv - \B\bv)^\top (\uv - \B\bv)
    + \lambda \mathcal{J}(\bv; d),
\end{equation}
where $\lambda$ is the smoothing parameter that governs the trade-off between
smoothness and closeness to the data. We use a quadratic difference penalty of
order $d$ on the coefficients , i.e., $\mathcal{J}(\bv; d) = \sum_j (\Delta^d
\beta_j)^2$, with $\Delta^1 \beta_j = \Delta \beta_j := \beta_j - \beta_{j -1}$.
By applying the $\Delta$ operator recursively we get $\Delta^2 \beta_j =
\Delta(\Delta \beta_j) = \beta_j - 2\beta_{j-1} + \beta_{j -2}$, etc. In matrix
notation the penalty can be written as
\begin{equation}\label{eq:pen_pspline}
  \mathcal{J}(\bv; d) = \bv^\top \D_{(d)}^\top \D_{(d)} \bv.
\end{equation}
The difference matrices $\D_{(d)}$ are constructed such that they lead to the
appropriate differences: first- and second-order differences result from
matrices of the form
\begin{equation}\label{eq:diffmat1}
  \D_{(1)}=\begin{pmatrix}
    -1     &      1 &        & \\
           & \ddots & \ddots & \\
           &        &     -1 & 1\\
         \end{pmatrix}
\end{equation}
and
\begin{equation}\label{eq:diffmat2}
    \D_{(2)}=\begin{pmatrix}
         1 &     -2 &      1 &        & \\
           & \ddots & \ddots &  \ddots & \\
           &        &      1 &  -2     & 1\\
  \end{pmatrix},
\end{equation}
where empty cells are equal to zero. Higher order difference penalties can be
easily derived. Difference penalties of order one penalize the deviation from a
constant. Second-order differences penalize the deviation from a straight line.
In general, differences of order $d$ penalize deviations from polynomials of
order $d-1$. The unpenalized effects, i.e., the constant (d = 1) or the straight
line (d = 2) are called the null space of the penalty. The null space remains
unpenalized, even in the limit of $\lambda \rightarrow \infty$.

The penalized least-squares criterion~\eqref{eq:pen_LS_pspline} is optimized in
each boosting step irrespective of the underlying distribution assumption. The
distribution assumption, or more generally, a specific loss function, is only
used to derive the appropriate negative gradient of the loss function.To fit the
negative gradient vector $\uv$, we fix the smoothing parameter $\lambda$
separately for each base-learner such that the corresponding degrees of freedom
of the base-learner are relatively small (typically not more than 4 to 6 degrees
of freedom). The boosting algorithm iteratively updates one base-learner per
boosting iteration. As the same base-learner can enter the model multiple times,
the final effect, which is the sum of all effect estimates for this
base-learner, can adapt to arbitrarily higher order smoothness. For details see
\citet{buehlmann03} and \citet{Hofner:unbiased:2011}.

\subsection{Estimating Cyclic Smooth Effects}
\label{sec:cyclic-p-splines}

P-splines with a cyclic constraint \citep{EilersMarx:2010} can be used to model
periodic, seasonal data. The cyclic B-spline basis functions are constructed
without knot expansion (Figure~\ref{fig:cyclic}). The B-splines are ``wrapped''
at the boundary knots. The boundary knots $\xi_0$ and $\xi_J$ (equal to
$\xi_{12}$ in Fig.~\ref{fig:cyclic}) play a central role in this setting as they
specify the points where the function estimate should be smoothly joined. If $x$
is, for example, the time during the day, then $\xi_0$ is 0:00, whereas $\xi_J$
is 24:00. Defining the B-spline basis in this fashion leads to a cyclic B-spline
basis with the $(n \times (J + 1))$ design matrix $\B_{\text{cyclic}}$. The
corresponding coefficients are collected in the $((J + 1) \times 1)$ vector $\bv
= (\beta_0, \ldots, \beta_J)$.

\begingroup
\setkeys{Gin}{width=1\hsize}

\begin{figure}
  \centering
\begingroup
\includegraphics{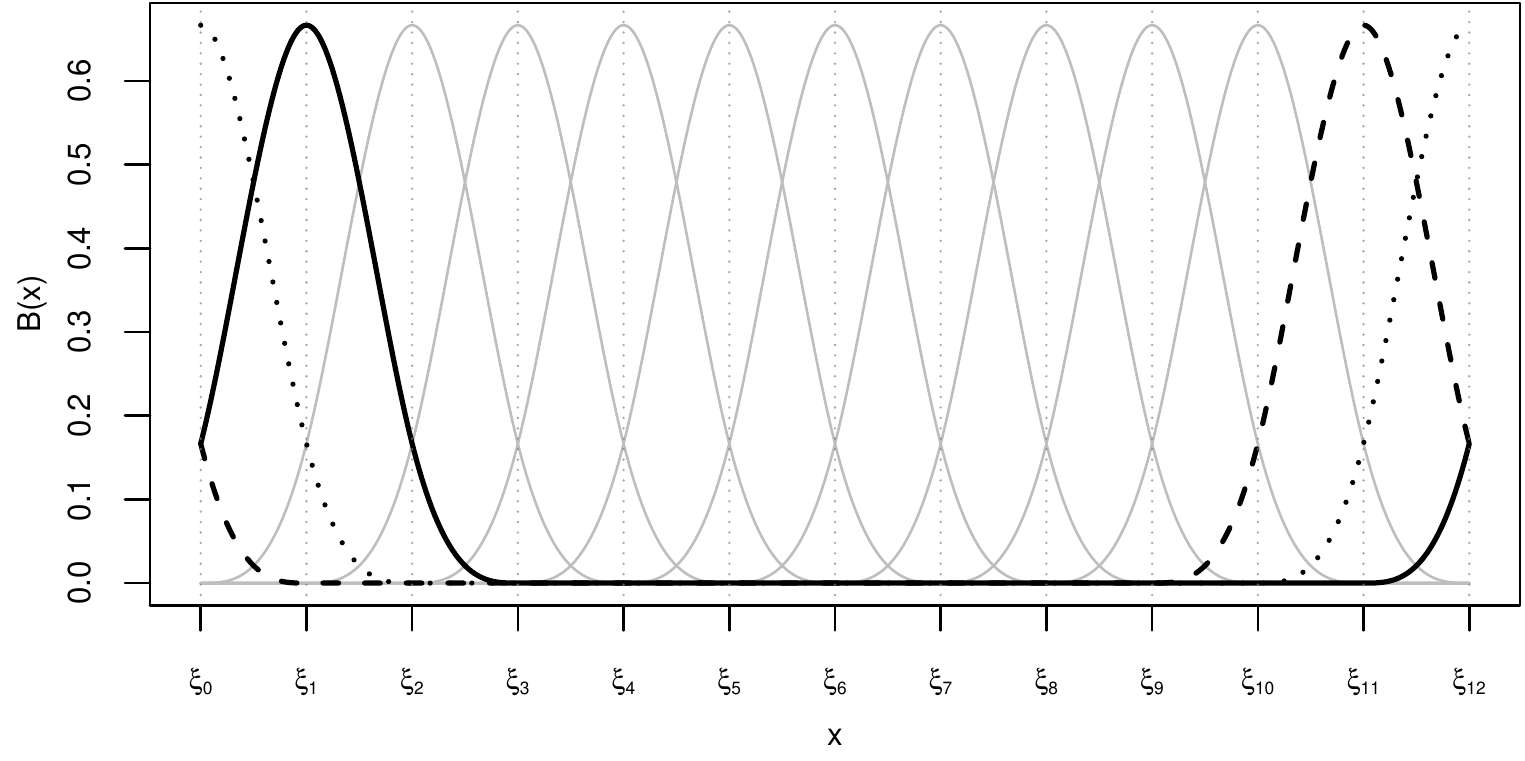}
\caption{Illustration of cyclic P-splines of degree three, with 11 inner knots
  and boundary knots $\xi_0$ and $\xi_{12}$. The gray curves correspond to
  B-splines. The black curves correspond to B-splines that are ``wrapped'' at
  the boundary knots, leading to a cyclic representation of the function. The
  dashed B-spline basis depends on observations in $[\xi_9, \xi_{12}] \cup
  [\xi_0, \xi_1]$ and thus on observations from both ends of the range of the
  covariate. The same holds for the other two black B-spline curves (solid and
  dotted).}
  \label{fig:cyclic}
\endgroup
\end{figure}

\endgroup

Specifying a cyclic basis guarantees that the resulting function estimate is
continuous in the boundary knots. However, no smoothness constraint is imposed
so far. This can be achieved by a cyclic difference penalty, for example,
$\mathcal{J}_{\text{cyclic}}(\mbf{\beta}) = \sum_{j = 0}^J (\beta_j -
\beta_{j-1})^2$ (with $d = 1$) or $\mathcal{J}_{\text{cyclic}}(\mbf{\beta}) =
\sum_{j = 0}^J (\beta_j - 2 \beta_{j-1} + \beta_{j - 2})^2$ (with $d = 2$),
where the index $j$ is ``wrapped'', i.e., $j := J + 1 + j$ if $j < 0$. Thus, the
differences between $\beta_0$ and $\beta_J$ or even $\beta_{J - 1}$ are taken
into account for the penalty. Hence, the boundaries of the support are
stabilized, and smoothness in and around the boundary knots is enforced. This
can also be seen in Figure~\ref{fig:cyclic_est}. The non-cyclic estimate
(Fig.~\ref{fig:cyclic_est_noncycl}) is less stable at the boundaries. As a
consequence, the ends do not meet. The cyclic estimate
(Fig.~\ref{fig:cyclic_est_cycl}), in contrast, is stabilized at the boundaries,
and the ends are smoothly joined.

\begingroup
\setkeys{Gin}{width=0.4\textwidth}

\begin{figure}\sidecaption
  \centering
\begingroup
\subfigure{(a)\label{fig:cyclic_est_noncycl}
  \raisebox{-\height}{
\includegraphics{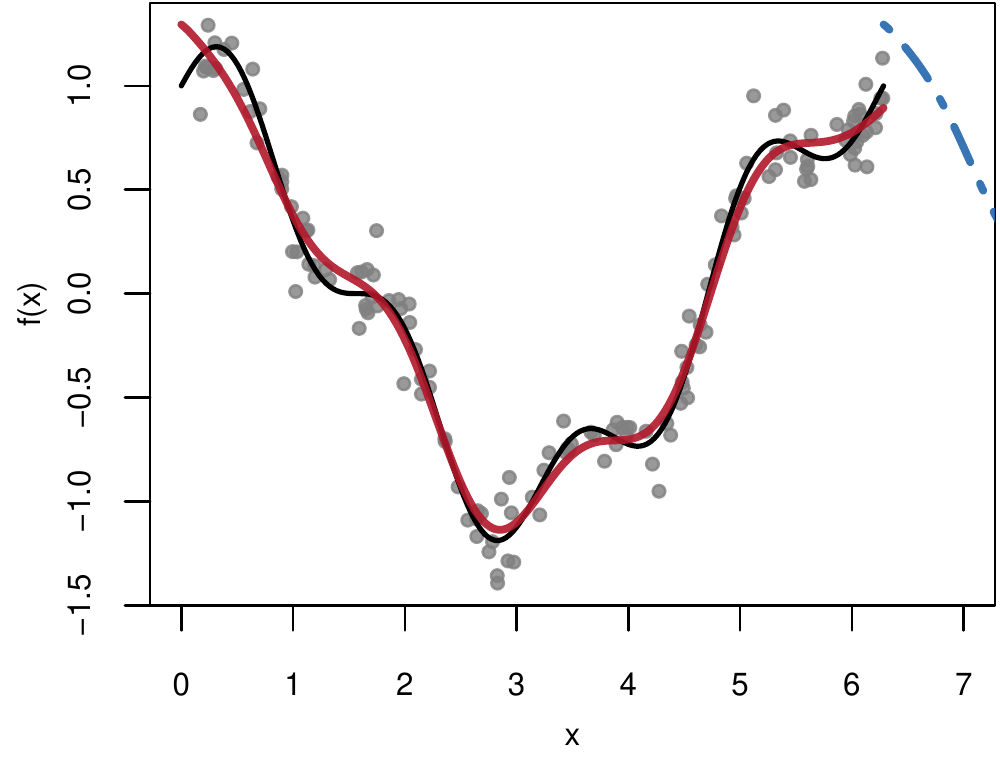}
}}
\hspace{1cm}
\subfigure{(b)\label{fig:cyclic_est_cycl}
  \raisebox{-\height}{
\includegraphics{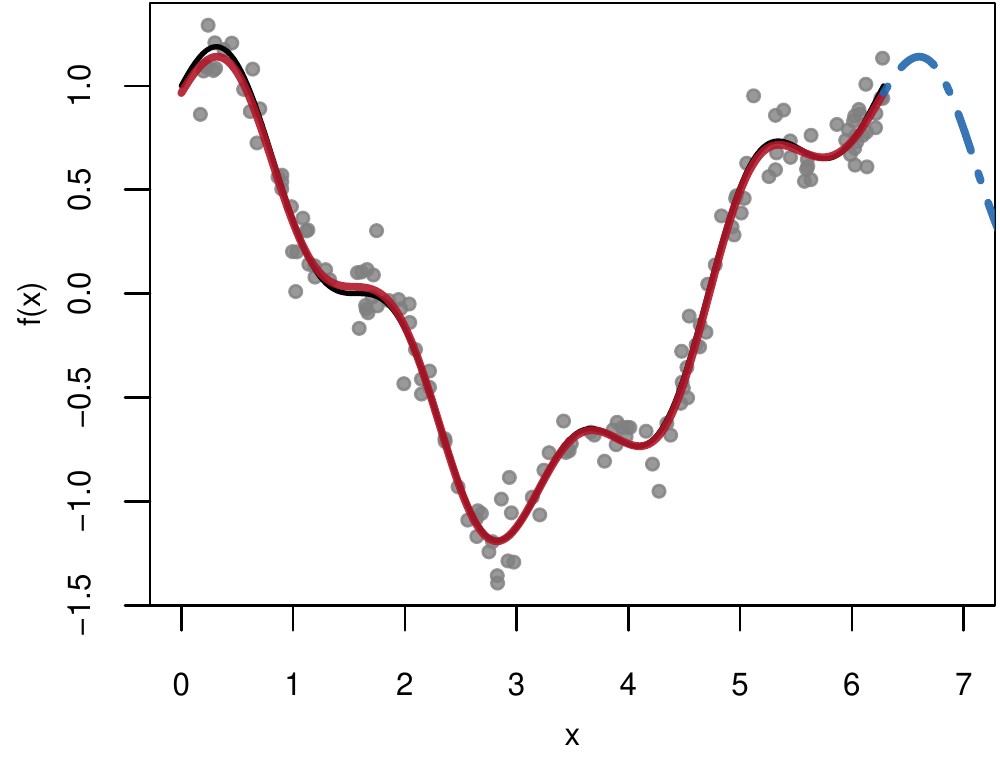}
}}
\endgroup
\caption{\subref{fig:cyclic_est_noncycl} Non-cyclic and
  \subref{fig:cyclic_est_cycl} cyclic P-splines. The red curve is the function
  estimated P-spline function. The blue, dashed curve is the same function
  shifted one period to the right. The data were simulated from a cyclic
  function with period $2 \pi$: $f(x) = \cos(x) + 0.25 \sin(4x)$ (black line)
  and realizations with additional normally distributed errors ($\sigma = 0.1$;
  gray dots). The cyclic estimate is closer to the true function and more stable
  in the boundary regions, and the ends meet (see \subref{fig:cyclic_est_cycl}
  at $x = 2\pi$).}
  \label{fig:cyclic_est}
\end{figure}

\endgroup

In matrix notation the penalty can be written as
\begin{equation}
  \mathcal{J}_{\text{cyclic}}(\bv; d) = \bv^\top \widetilde{\D}^\top_{(d)} \widetilde{\D}_{(d)}\bv,
\end{equation}
with difference matrices
\begin{equation}\label{eq:cyclic_diff1}
  \widetilde{\D}_{(1)} =
  \begin{pmatrix}
     1 &     &     &  -1 \\
    -1 &   1 &     &     \\
       &  \ddots &   \ddots &     \\
       &     &  -1 &   1 \\
  \end{pmatrix}
\end{equation}
and
\begin{equation}\label{eq:cyclic_diff2}
  \widetilde{\D}_{(2)} =
  \begin{pmatrix}
    1 &      &     &  1  & -2 \\
   -2 &   1  &     &     &  1 \\
    1 &  -2  &  1  &     &    \\
      &  \ddots & \ddots & \ddots  & \\
      &     &  1  & -2  &  1 \\
  \end{pmatrix},
\end{equation}
where empty cells are equal to zero. Coefficients can then be estimated using
the penalized least-squares criterion~(\ref{eq:pen_LS_pspline}), where the
design matrix and the penalty matrix are replaced with the corresponding cyclic
counterparts, i.e., $\mathcal{Q}(\bv) = (\uv - \B_{\text{cyclic}}\,\bv)^\top
(\uv - \B_{\text{cyclic}}\,\bv) + \lambda \mathcal{J}_{\text{cyclic}}(\bv; d)$.
Again, we fix the smoothing parameter $\lambda$ and control the smoothness of
the final fit by the number of boosting iterations.

As mentioned in Section~\ref{sec:p-splines}, P-splines have a null space, i.e.,
an unpenalized effect, which depends on the order of the differences in the
penalty. However, cyclic P-splines have a null space that includes only a
constant, irrespective of the order of the difference penalty. Globally seen,
i.e., for the complete function estimate, the order of the penalty plays no role
(even in the limit $\lambda \rightarrow \infty$). Locally, however, the order of
the difference penalty has an effect. For example, with $d = 2$, the estimated
function is penalized for deviations from linearity and hence, locally
approaches a straight line (with increasing $\lambda$).

An empirical evaluation of cyclic P-splines shows the clear superiority of
cyclic splines compared to unconstrained estimates, both with respect to the MSE
and the conformance with the cyclicity assumption
\citepappendix{sec:sim_cyclic-effects}.

\subsection{Estimating Monotonic Effects}
\label{sec:monotonic-effects}

To achieve a smooth, yet monotonic function estimate, \citet{Eilers2005}
introduced P-splines with an \emph{additional asymmetric difference penalty}.
The penalized least-squares criterion \eqref{eq:pen_LS_pspline} becomes
\begin{equation}\label{eq:constrained_pen_ls}
  \begin{split}
  \mathcal{Q}(\bv) = &(\uv - \B\bv)^\top (\uv - \B\bv)
    + \lambda_1 \mathcal{J}(\bv; d)
    \\ &+ \lambda_2 \mathcal{J}_{\text{asym}}(\bv; c),
  \end{split}
\end{equation}
with the quadratic difference penalty $\mathcal{J}(\bv; d)$ as in standard
P-splines \Eqref{eq:pen_pspline} and an additional asymmetric difference penalty
of order $c$
\begin{equation}\label{eq:pen_asym}
  \mathcal{J}_{\text{asym}}(\bv; c) = \sum_{j = c + 1}^J v_j (\Delta^c \beta_j)^2
                                = \bv^\top \D_{(c)}^\top \V \D_{(c)} \bv,
\end{equation}
where the difference matrix $\D_{(c)}$ is constructed as in
Equations~\eqref{eq:diffmat1} and \eqref{eq:diffmat2}. This asymmetric
difference penalty ensures that the differences (of order $c$) of adjacent
coefficients are positive or negative. The choice of $c$ implies the type of the
additional constraint: monotonicity for $c = 1$ or convexity/concavity for $c =
2$. In the remainder of this article, we restrict our attention to monotonicity
constraints; however, one can also consider concave constraints. The asymmetric
penalty looks very much like the P-spline penalty~(\ref{eq:pen_pspline}) with
the important distinction of weights $v_j$, which are specified as
\begin{equation}\label{eq:weights}
  v_j = \begin{cases}
    0 & \text{if \, $\Delta^c \beta_j > 0$} \\
    1 & \text{if \, $\Delta^c \beta_j \leq 0$}.
  \end{cases}
\end{equation}
The weights are collected in the diagonal matrix $\V = \diag(\vv)$. With $c =
1$, this enforces monotonically \emph{increasing} functions. Changing the
direction of the inequalities in the distinction of cases leads to monotonically
\emph{decreasing} functions. Note that positive/negative differences of adjacent
coefficients are sufficient but not necessary for monotonically
increasing/decreasing effects. As the weights \eqref{eq:weights} depend on the
coefficients $\bv$, a solution to \eqref{eq:constrained_pen_ls} can only be
found by iteratively minimizing $\mathcal{Q}(\bv)$ with respect to $\bv$, where
the weights $\vv$ are updated in each iteration. The estimation converges if no
further changes in the weight matrix $\V$ occur. The penalty parameter
$\lambda_2$, which is associated with the additional
constraint~\eqref{eq:pen_asym}, should be chosen quite large \citep[e.g.,
$10^6$;][]{Eilers2005} and resembles the researcher's \emph{a priori} assumption
of monotonicity. Larger values are associated with a stronger impact of the
monotonic constraint on the estimation. The penalty parameter $\lambda_1$
associated with the smoothness constraint is usually fixed so that the overall
degrees of freedom of the smooth monotonic effect resemble a pre-specified
value. Again, by updating the base-learner multiple times, the effect can adapt
greater flexibility while keeping the monotonicity assumption. A detailed
discussion of monotonic P-splines in the context of boosting models is given in
\citet{Hofner:monotonic:2011}. In their presented framework, the authors also
derive an asymmetric difference penalty for monotonicity-constrained, ordered
categorical effects.

One can also use asymmetric difference penalties on differences of order 0,
i.e., on the coefficients themselves, to achieve smooth positive or negative
effect estimates. This idea can also be used to fit smooth effect estimates with
an arbitrarily fixed co-domain by specifying either upper or lower bounds or
both bounds at the same time.

\paragraph{Improved Fitting Method for Monotonic Effects}

An alternative to iteratively minimizing the penalized least-squares
criterion~\eqref{eq:constrained_pen_ls} to obtain smooth monotonic estimates, is
given by quadratic programming methods \citep{Goldfarb:1982,Goldfarb:1983}. To
fit the monotonic base-learner to the negative gradient vector $\uv$ using
quadratic programming, we minimize the penalized least-squares criterion
\eqref{eq:pen_LS_pspline} with the additional constraint
\begin{equation*} 
  \D_{(c)} \bv \geq \mbf{0},
\end{equation*}
with difference matrix $\D_{(c)}$ as defined above and null vector $\mbf{0}$ (of
appropriate dimension). To change the direction of the constraint, e.g., to
obtain monotonically decreasing functions, one can use the negative difference
matrix $-\D_{(c)}$. The results obtained by quadratic programming are
(virtually) identical to the results obtained by iteratively
solving~\eqref{eq:constrained_pen_ls} \tabappendix{tab:sim_res_iter}, but the
computation time can be greatly reduced. An empirical evaluation of monotonic
splines (fitted using quadratic programming methods) shows the superiority of
monotonic splines compared to unconstrained estimates: The MSE is comparable to
the MSE of unconstrained effects and a clear superiority is given with respect
to the conformance with the monotonicity assumption
\citepappendix{sec:sim_cyclic-effects}.

\subsection{Estimating Effects with Boundary Constraints}
\label{sec:boundary_constraints}

In some cases, e.g., for extrapolation, it might be of interest to impose
boundary constraints, such as constant or linear boundaries, to higher order
splines. These constraints can be enforced by using a strong penalty on, e.g.,
the three outer spline coefficients on each side of the range of the data or on
one side only. Constant boundaries are obtained by a strong penalty on the
first-order differences, while a strong second-order difference penalty results
in linear boundaries. Technically, this can be obtained by an additional penalty
\begin{equation}\label{eq:pen_3}
  \begin{split}
    \mathcal{J}_{\text{boundary}}(\bv; e)  &= \sum_{j = c + 1}^J v_j (\Delta^e
    \beta_j)^2\\ &= \bv^{\top} \D_{(e)}^{\top}\V^{(3)} \D_{(e)} \bv,
  \end{split}
\end{equation}
where $\D_{(e)}$ is a difference matrix of order $e$ (cf.\ Eq.~\eqref{eq:diffmat1}
and~\eqref{eq:diffmat2}). The weight $v^{(3)}_j$ is one if the corresponding
coefficient is subject to a boundary constraint. Thus, here the first and the
last three elements of $\vv^{(3)}$ are equal to one, and the remaining weights
are equal to zero. The weight matrix $\V^{(3)} = \diag(\vv^{(3)})$. Boundary
constraints can be successfully imposed on P-splines as well as on monotonic
P-splines by adding the penalty~\eqref{eq:pen_3} to the respective penalized
least-squares criterion. A quite large penalty parameter $\lambda_3$ associated
with the boundary constraint is chosen (e.g., $10^6$). For an application of
modeling the gas flow in gas transmission networks using monotonic effects with
boundary constraints, see \citet{Sobotka:Gasflow:2013}.

\subsection{Varying Coefficients}

Varying-coefficient models allow one to model flexible interactions in which the
regression coefficients of a predictor vary smoothly with one or more other
variables, the so-called effect modifiers \citep{Hast:Tibs:var-coef:1993}. The
varying coefficient term can be written as $f(x, z) = x \cdot \beta(z)$, where
$z$ is the effect modifier and $\beta(\cdot)$ a smooth function of $z$.
Technically, varying coefficients can be modeled by fitting the interaction of
$x$ and a basis expansion of $z$. Thus, we can use all discussed spline types,
such as simple P-splines, monotonic splines, or cyclic splines, to model
$\beta(z)$. Furthermore, bivariate P-splines as discussed in the following
section can be facilitated as well.

\section{Constrained Effects for Bivariate P-Splines}
\label{sec:constr-effects-bivar}

\subsection{Bivariate P-spline Base-learners}\label{sec:bivariate-p-splines}

Bivariate, or tensor product, P-splines are an extension of univariate P-splines
that allow modeling of smooth effects of two variables. These can be used to
model smooth interaction surfaces, most prominently spatial effects. A bivariate
B-spline of degree $\delta$ for two variables $x_1$ and $x_2$ can be constructed
as the product of two univariate B-spline bases $B_{jk}(x_1, x_2; \delta) =
B^{(1)}_j(x_1; \delta) \cdot B^{(2)}_k(x_2; \delta)$. The bivariate B-spline
basis is formed by all possible products $B_{jk}$, $j = 1, \ldots, J$, $k = 1,
\ldots, K$. Theoretically, different numbers of knots for $x_1$ ($J$) and $x_2$
($K$) are possible, as well as B-spline basis functions with different degrees
$\delta_1$ and $\delta_2$ for $x_1$ and $x_2$, respectively. A bivariate
function $f(x_1, x_2)$ can be approximated as $ f(x_1, x_2) = \sum_{j = 1}^J
\sum_{k = 1}^K \beta_{jk} B_{jk}(x_1, x_2) = \B(x_1, x_2)^\top \bv,$ where the
vector of B-spline bases for variables $(x_1, x_2)$ equals $\B(x_1, x_2) =
\Bigl(B_{11}(x_1, x_2), \ldots,\allowbreak B_{1K}(x_1, x_2),\allowbreak
B_{21}(x_1, x_2), \ldots, B_{JK}(x_1, x_2) \Bigr)^\top,$ and the coefficient
vector
\begin{equation}\label{eq:spatial_coefs}
  \bv = (\beta_{11}, \ldots, \beta_{1K}, \beta_{21}, \ldots, \beta_{JK})^\top.
\end{equation}
The $(n \times JK)$ design matrix then combines the vectors $\B(\x_i)$ for
observations $\x_{i} = (x_{i1}, x_{i2}),\, i = 1, \ldots, n,$ such that the
$i$th row contains $\B(\x_i)$, i.e., $\B = \Bigl(\B(\x_{1}), \ldots, \B(\x_{i}),
\ldots, \B(\x_{n}) \Bigr)^\top$. The design matrix $\B$ can be conveniently
obtained by first evaluating the univariate B-spline bases $\B^{(1)} =
\Bigl(B^{(1)}_j(x_{i1})\Bigr)_{i = 1, \ldots, n \atop j = 1, \ldots, J}$ and
$\B^{(2)} = \Bigl(B^{(2)}_k (x_{i2})\Bigr)_{i = 1, \ldots, n \atop k = 1, \ldots,
  K}$ of the variables $x_1$ and $x_2$ and subsequently constructing the design
matrix as
\begin{equation}\label{eq:designmat_kronecker}
  \B = (\B^{(1)} \otimes \mbf{e}^\top_K) \odot (\mbf{e}^\top_J \otimes \B^{(2)}),
\end{equation}
where $\mbf{e}_K = (1, \ldots, 1)^\top$ is a vector of length $K$ and $\mbf{e}_J
= (1, \ldots, 1)^\top$ a vector of length $J$. The symbol $\otimes$ denotes the
Kronecker product and $\odot$ denotes the element-wise product. Definitions of
both products are given in \citetappendix{sec:matrix-algebra}.

As for univariate P-splines, a suitable penalty matrix is required to enforce
smoothness. The bivariate penalty matrix can be constructed from separate,
univariate difference penalties for $x_1$ and $x_2$, respectively. Consider the
$(J \times J)$ penalty matrix $\K^{(1)} = (\D^{(1)})^\top \D^{(1)}$ for $x_1$, and the $(K
\times K)$ penalty matrix $\K^{(2)} = (\D^{(2)})^\top \D^{(2)}$ for $x_2$. The penalties
are constructed using difference matrices $\D^{(1)}$ and $\D^{(2)}$ of (the same) order
$d$. However, different orders of differences $d_1$ and $d_2$ could be used if
this is required by the data at hand. The combined difference penalty can then
be written as the sum of Kronecker products
\begin{equation}\label{eq:spatial_pen}
  \mathcal{J}_{\text{tensor}}(\bv; d) = \bv^\top (\K^{(1)} \otimes \I_K + \I_J
  \otimes \K^{(2)}) \bv,
\end{equation}
with identity matrices $\I_J$ and $\I_K$ of dimension $J$ and $K$, respectively.

With the negative gradient vector $\uv$ as response, models can then be
estimated by optimizing the penalized least-squares criterion in analogy to
univariate P-splines:
\begin{equation}\label{eq:spatial_PLS}
  \mathcal{Q}(\bv) = (\uv - \B\bv)^\top (\uv - \B\bv)
    + \lambda \mathcal{J}_{\text{tensor}}(\bv; d),
\end{equation}
with design matrix~\eqref{eq:designmat_kronecker},
penalty~\eqref{eq:spatial_pen} and fixed smoothing parameter $\lambda$. For more
details on tensor product splines, we refer the reader to \citet{Wood2006}.
\citet{Kneib:Hothorn:Tutz:modelchoice:2009} give an introduction to tensor
product P-splines in the context of boosting.

\subsection{Estimating Bivariate Cyclic Smooth Effects}
\label{sec:bivar-cyclic}

Based on bivariate P-splines, cyclic constraints in both directions of $x_1$ and
$x_2$ can be straightforwardly implemented. One builds the univariate, \emph{cyclic}
design matrices $\B_{\text{cyclic}}^{(1)}$ and $\B_{\text{cyclic}}^{(2)}$ for
$x_1$ and $x_2$, respectively. The bivariate design matrix then is
$\B_{\text{cyclic}} = (\B_{\text{cyclic}}^{(1)} \otimes \mbf{e}^\top_K) \odot
(\mbf{e}^\top_J \otimes \B_{\text{cyclic}}^{(2)})$, as in
Equation~\eqref{eq:designmat_kronecker}.

With the univariate, \emph{cyclic} difference matrices $\widetilde{\D}^{(1)}$
for $x_1$ and $\widetilde{\D}^{(2)}$ for $x_2$ (cf.\ Eq.~\eqref{eq:cyclic_diff1}
and~\eqref{eq:cyclic_diff2}), we obtain cyclic penalty matrices $\K^{(1)} =
(\widetilde{\D}^{(1)})^\top \widetilde{\D}^{(1)}$ and $\K^{(2)} =
(\widetilde{\D}^{(2)})^\top \widetilde{\D}^{(2)}$. Thus, in analogy to the usual
bivariate P-spline penalty~\Eqref{eq:spatial_pen}, the bivariate cyclic penalty
can be written as $\mathcal{J}_{\text{cyclic, tensor}}(\bv) = \bv^\top (\K^{(1)}
\otimes \I_K + \I_J \otimes \K^{(2)}) \bv$, in analogy to
Equation~\eqref{eq:spatial_coefs}. Estimation is then a straightforward
application of the penalized least-squares criterion as
in~\Eqref{eq:spatial_PLS} with cyclic design and penalty matrices. An example of
bivariate, cyclic splines is given in Section~\ref{sec:roe-deer-activity}, where
a cyclic surface is used to estimate the combined effect of time (during the
day) and calendar day on roe deer activity.

\subsection{Estimating Bivariate Monotonic Effects}
\label{sec:bivar-mono}

As shown in \citet{Hofner:Dissertation:2011}, it is not sufficient to add
monotonicity constraints to the marginal effects because the resulting
interaction surface might still be non-monotonic. Thus, instead of the marginal
functions, the complete surface needs to be constrained in order to achieve a
monotonic surface. Therefore, we utilized bivariate P-splines and added
monotonic constraints for the row- and column-wise differences of the matrix of
coefficients $\mathcal{B} = \Bigl( \beta_{jk} \Bigr)_{j = 1, \ldots, J;\;\; k =
  1, \ldots, K}$. As proposed by \citet{Bollaerts_L2_2006}, one can use two
independent asymmetric penalties to allow different directions of
monotonicity, i.e., increasing in one variable, e.g., $x_1$, and decreasing in
the other variable, e.g., $x_2$, or one can use different prior assumptions of
monotonicity reflected in different penalty parameters $\lambda$. Let $\B$
denote the $(n \times JK)$ design matrix \eqref{eq:designmat_kronecker}
comprising the bivariate B-spline bases of $\x_i$, and let $\bv$ denote the
corresponding $(JK \times 1)$ coefficient vector \eqref{eq:spatial_coefs}.
Monotonicity is enforced by the asymmetric difference penalties
\begin{align*}
  \mathcal{J}_{\text{asym},1}(\bv; c) &=
          \sum_{j = c + 1}^J \sum_{k = 1}^K v^{(1)}_{jk} (\Delta^c_1 \beta_{jk})^2\\ &=
          \bv^\top (\D^{(1)} \otimes \I_K)^\top \V^{(1)} (\D^{(1)} \otimes \I_K) \bv,\\
  \mathcal{J}_{\text{asym},2}(\bv; c) &=
          \sum_{j = 1}^J \sum_{k = c + 1}^K v^{(2)}_{jk} (\Delta^c_2 \beta_{jk})^2\\ &=
          \bv^\top (\I_J \otimes \D^{(2)})^\top \V^{(2)} (\I_J \otimes \D^{(2)})
          \bv,
\end{align*}
where $\Delta^c_1$ are the column-wise and $\Delta^c_2$ the row-wise differences
of order $c$, i.e., $\Delta^1_1 \beta_{jk} = \beta_{jk} - \beta_{(j-1)k}$ and
$\Delta^1_2 \beta_{jk} = \beta_{jk} - \beta_{j(k-1)}$, etc. Thus,
$\mathcal{J}_{\text{asym},1}$ is associated with constraints in the direction of
$x_1$, while $\mathcal{J}_{\text{asym},2}$ acts in the direction of $x_2$. The
corresponding difference matrices are denoted by $\D^{(1)}$ and $\D^{(2)}$. The
weights $v^{(l)}_{jk},\, l = 1, 2$ are specified in analogy to
\eqref{eq:weights}, i.e., with $c = 1$, we obtain monotonically increasing
estimates with weights $v^{(l)}_{jk} = 1$ if $\Delta_l^c \beta_{jk} \leq 0$, and
$v^{(l)}_{jk} = 0$ otherwise. Changing the inequality sign leads to
monotonically decreasing function estimates. Differences of order $c = 2$ lead
to convex or concave constraints. For the matrix notation, the weights are
collected in the diagonal matrices $\V^{(l)} = \diag(\vv^{(l)})$. The constraint
estimation problem for monotonic surface estimates in matrix notation becomes
\begin{equation}
\begin{split}\label{eq:constrained_pen_ls_spatial}
  \mathcal{Q}(\bv) = &(\uv - \B\bv)^\top (\uv - \B\bv)
    + \lambda_1 \mathcal{J}_{\text{tensor}}(\bv; d) \\
    &+ \lambda_{21} \mathcal{J}_{\text{asym},1}(\bv; c)
    + \lambda_{22} \mathcal{J}_{\text{asym},2}(\bv; c),
\end{split}
\end{equation}
where $\mathcal{J}_{\text{tensor}}(\bv; d)$ is the standard bivariate P-spline
penalty of order $d$ with corresponding fixed penalty parameter $\lambda_1$. The
penalty parameters $\lambda_{21}$ and $\lambda_{22}$ are associated with
constraints in the direction of $x_1$ and $x_2$, respectively. To enforce
monotonicity in both directions, one should choose relatively large values for
both penalty parameters (e.g., $10^6$). Setting either of the two penalty
parameters to zero results in an unconstrained estimate in this direction, with
a constraint in the other direction. For example, by setting $\lambda_{21} = 0$
and $\lambda_{22} = 10^6$, one gets a surface that is monotone in $x_2$ for each
value of $x_1$ but is not necessarily monotone in $x_1$.

\paragraph{Model Fitting for Monotonic Base-Learners}

Model estimation can be achieved by using either the iterative algorithm to
solve Equation~\eqref{eq:constrained_pen_ls_spatial} or quadratic programming
methods as in the univariate case described in
Section~\ref{sec:monotonic-effects}. In the latter case, we minimize the
penalized least-squares criterion~\eqref{eq:spatial_PLS} subject to the
constraints $(\D^{(1)} \otimes \I_K) \bv \geq \mbf{0}$ and $(\I_J \otimes
\D^{(2)}) \geq \mbf{0}$, i.e., we constrain the row-wise or column-wise
differences to be non-negative. As above, multiplying the difference matrices by
$-1$ leads to monotonically decreasing estimates. Using the two constraints is
equivalent to requiring
\begin{equation*}
  \begin{pmatrix}
    \D^{(1)} \otimes \I_K \\
    \I_J \otimes \D^{(2)}
  \end{pmatrix} \bv \geq \mbf{0}.
\end{equation*}

\section[Confidence Intervals and Confidence Bands]{Confidence Intervals and Confidence Bands}
\label{sec:CIs}

In general it is difficult to obtain theoretical confidence bands for penalized
regression models in a frequentist setting. In the boosting context, we select
the best fitting base-learner in each iteration and additionally shrink the
parameter estimates. To reflect both, the shrinkage and the selection process,
it is necessary to use bootstrap methods in order to obtain confidence
intervals. Based on the bootstrap one can draw random samples from the empirical
distribution of the data, which can be used to compute empirical confidence
intervals based on point-wise quantiles of the estimated functions
\citep{Hofner:BoostingSurvival:2013,Schmid:BoostedBeta:2013}. The optimal
stopping iteration should be obtained within each of these bootstrap samples,
i.e., a nested bootstrap is advised. We propose to use 1000 outer bootstrap
samples for the confidence intervals. We will give examples in the case studies
below.

To obtain simultaneous $P\%$ confidence bands, one can use the point-wise
confidence intervals and rescale these until $P\%$ of all curves lie within
these bands \citep{Krivobokova:2010:simultaneous}. An alternative to confidence
intervals and confidence bands is stability selection
\citep{Meinshausen:2010,ShahSamworth:Stabsel:2013,Hofner:stabsel:2014}, which
adds an error control (of the per-family error rate) to the built-in selection
process of boosting.

\section[Computational Details]{Computational Details}
\label{sec:software}

The \textsf{R} system for statistical computing \citep{RCore:2014} was used to
implement the analyses. The package \textbf{mboost} 
\citep{Hothorn:mboost_MLOSS:2010,pkg:mboost:CRAN:2.4,Hofner:mboost:2014}
implements the newly developed framework for constrained regression models based
on boosting. Unconstrained additive models were fitted using the function
\R{gam()} from package \textbf{mgcv} \citep{Wood2006:GAM,pkg:mgcv}, and function
\R{scam()} from package \textbf{scam}
\citep{Pya:pkg:scam:2014,PyaWood:scam:2014} was used to fit constrained
regression models based on a Newton-Raphson method.

In \textbf{mboost}, one can use the function \texttt{gamboost()} to fit
structured additive models. Models are specified using a formula, where one can
define the base-learners on the right hand side: \texttt{bols()} implements
ordinary least-squares base-learners (i.e., linear effects), \texttt{bbs()}
implements P-spline base-learners, and \texttt{brandom()} implements random
effects. Constrained effects are implemented in \texttt{bmono()} (monotonic
effects, convex/concave effects and boundary constraints) and \texttt{bbs(...,
  cyclic = TRUE)} (cyclic P-spline base-learners). Confidence intervals for
boosted models can be obtained for fitted models using \texttt{confint()}. For
details on the usage see the \textsf{R} code, which is given as electronic
supplement.

\section{Case Studies}\label{sec:case-studies}

In order to demonstrate the wide range of applicability of the derived
framework, we show three case studies with different challenges, in the
following sections. These include a) the combination of monotonic and cyclic
effect estimates in the context of Poisson models, b) the estimation of a
bivariate cyclic effect and cyclic varying coefficients in a Gaussian model, and
c) the application of a tensor product of a monotonic and a cyclic effect to a
fit conditional distribution model which models the complete distribution at
once.

\subsection{S\~{a}o Paulo Air Pollution}\label{sec:sao-paulo}

In this study, we examined the effect of air pollution in S\~{a}o Paulo on
mortality. \citet{Saldiva:1995} investigated the impact of air pollution on
mortality caused by respiratory problems of elderly people (over 65 years of
age). We concentrated on the effect of SO$_2$ on mortality of elderly people. We
considered a Poisson model for the number of respiratory deaths of the form
\begin{align*}
  \log(\mbf{\mu}) = & \x^\top \mbf{\beta} + f_1(\text{day of
    the year}) + f_2(\text{time})\\ &+ f_3(\text{SO$_2$}),
\end{align*}
where the expected number of death due to respiratory causes $\mbf{\mu}$ is
related to a linear model with respect to covariates $\x$, such as temperature,
humidity, days of week, and non-respiratory deaths. Additionally, we wanted to
adjust for temporal changes; the study was conducted over four successive years
from January 1994 to December 1997. This allows us to decompose the temporal
effect into a smooth cyclicity-constrained effect for the day of the year
($f_1$, seasonal effect) and a smooth long-term trend for the variation over the
years ($f_2$). Finally, we added a smooth effect $f_3$ of the pollutant's
concentration. Smooth estimates of the effect of SO$_2$ on respiratory deaths
behaved erratically. This seems unreasonable as an increase in the air pollutant
should not result in a decreased risk of death. Hence, a monotonically
increasing effect should lead to a more stable model that can be interpreted.
Details of the data set can be found in \citetappendix{sec:sao_paulo_data},
where we also describe the model specification in detail.

In addition to our boosting approach (\mbmono{}), we used the \textbf{scam}
package \citep{Pya:pkg:scam:2014,PyaWood:scam:2014} to fit a shape constrained
model (\scam{}) with essentially the same model specifications as for the
boosting model \mbmono{}. We also fitted an unconstrained additive model using
\textbf{mgcv} \citep{Wood2006:GAM,pkg:mgcv} (\mgcv{}). In this model we did not
use the decomposition of the seasonal effect but fitted a single smooth effect
over time (that combines the seasonal pattern and the long term trend) and we
did not use a monotonicity constraint. Otherwise, the model specifications are
essentially the same as for the boosting model \mbmono{}. \textsf{R} code to fit
all discussed models is given as electronic supplement.

\paragraph{Results}

\begingroup
\setkeys{Gin}{width=0.27\textwidth}

\begingroup
\begin{figure*}
  \centering

\subfigure{(a)\label{fig:sao_paulo_results_a}
  \raisebox{-\height}{
\includegraphics{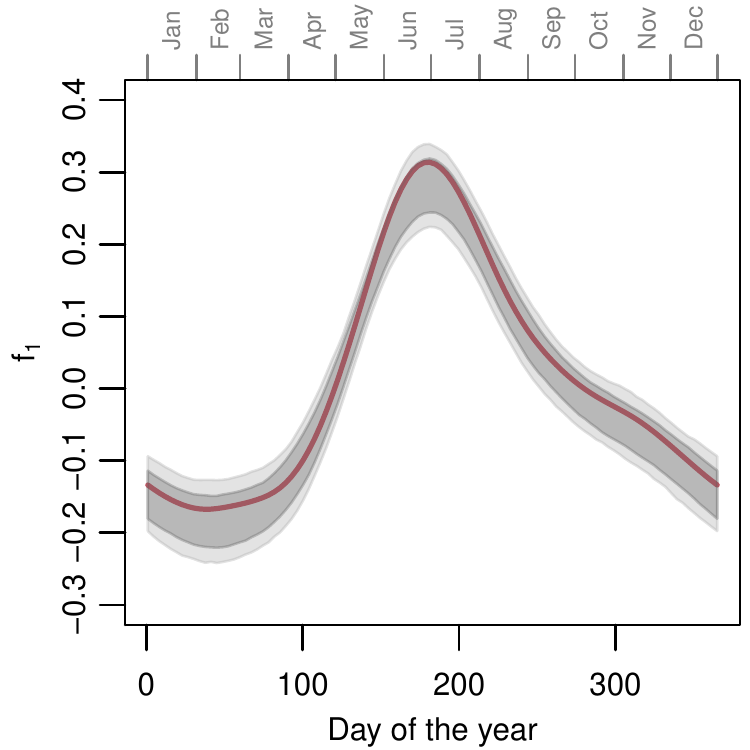}
}}
\hfill
\subfigure{(b)\label{fig:sao_paulo_results_b}
  \raisebox{-\height}{
\includegraphics{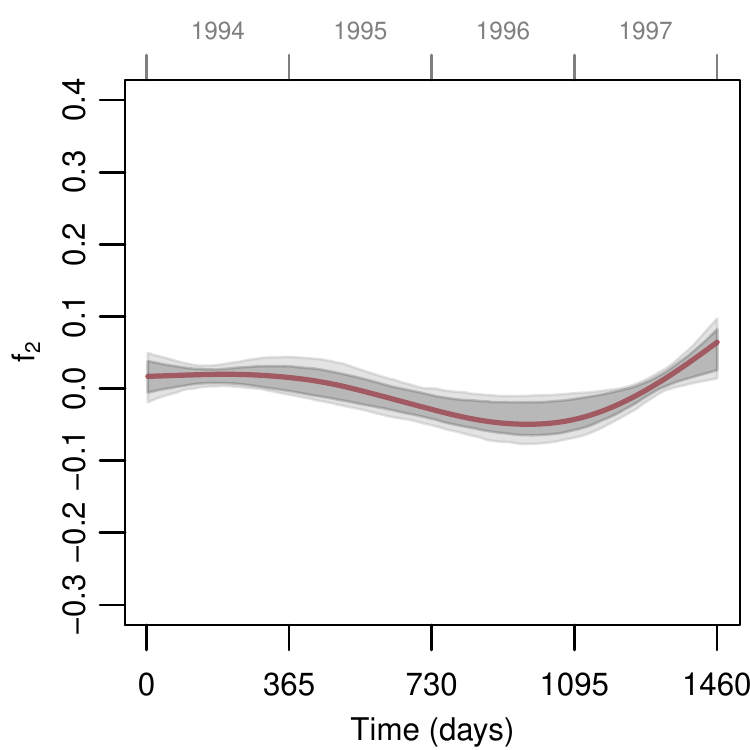}
}}
\hfill
\subfigure{(c)\label{fig:sao_paulo_results_c}
  \raisebox{-\height}{
\includegraphics{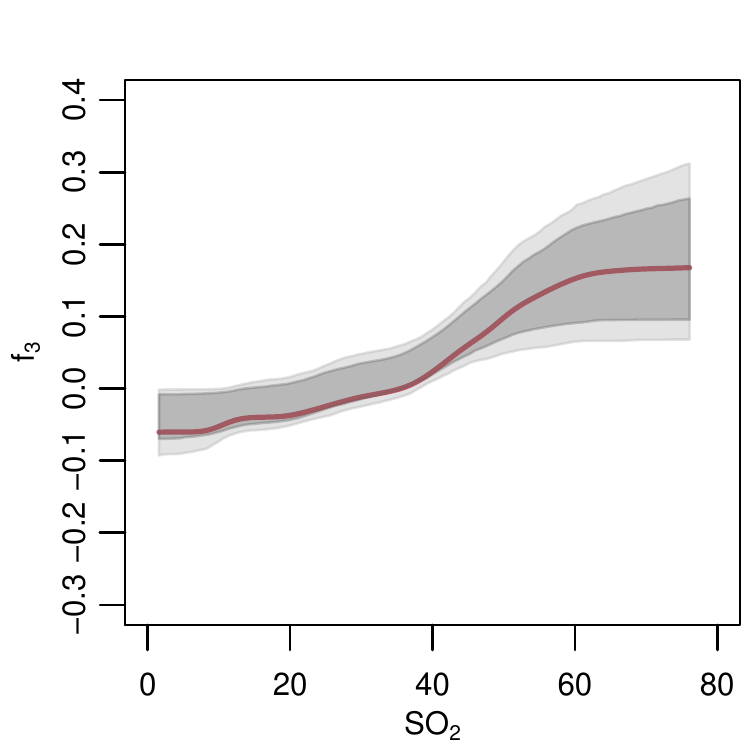}
}}
\caption{Estimated effects of the model \mbmono{} together with 80\% (dark gray)
  and 95\% (light gray) point-wise bootstrap confidence intervals:
  \subref{fig:sao_paulo_results_a} cyclic seasonal effect,
  \subref{fig:sao_paulo_results_b} unconstrained long-term trend, and
  \subref{fig:sao_paulo_results_c} monotonic effect of
  SO$_2$. \label{fig:sao_paulo_results}}
\end{figure*}

\endgroup

\endgroup

As we used a cyclic constraint for the seasonal effect, the ends of the function
estimate meet, i.e., day 365 and day 1 are smoothly joined (see
Figure~\ref{fig:sao_paulo_results_a}). The effect showed a clear peak in the
cool and dry winter months (May to August in the southern hemisphere) and a
decreased risk of mortality in the warm summer months. This is in line with the
results of other studies \citep[e.g.,][]{Saldiva:1995}. In the trend over the
years (Figure~\ref{fig:sao_paulo_results_b}), mortality decreased from 1994 to
1996 and increased thereafter. However, one should keep in mind that this trend
needs to be combined with the periodical effect to form the complete temporal
pattern.

The estimated smooth effect for the pollutant SO$_2$ resulting from the model
\mbmono{} (Figure~\ref{fig:sao_paulo_results_c}) indicated that an increase
of the pollutant's concentration does not result in a (substantially) higher
mortality up to a concentration of $40\,\mu$g/m$^3$. From this point onward, a
steep increase in the expected mortality was observed, which flattened again for
concentrations above $60\,\mu$g/m$^3$. Hence, a dose-response relationship was
observed, where higher pollutant concentrations result in a higher expected
mortality. At the same time, the model indicated that increasing pollutant
concentrations are almost harmless until a threshold is exceeded, and that the
harm of SO$_2$ is not further increased after reaching an upper threshold. In an
investigation of the effect of PM$_{10}$ \citet{Saldiva:1995} found no ``safe''
threshold in their study of elderly people in S\~{a}o Paulo. They also
investigated the effect of SO$_2$ but did not report on details, such as
possible threshold values, in this case. The more recent study on the effect of
air pollution in S\~{a}o Paulo on children \citep{Conceicao:2001} used only
linear effects for pollutant concentrations. Hence, no threshold values can be
estimated.

The linear effects of the model \mbmono{} (results not presented here; see
\textsf{R} code in the electronic supplement) showed (small) negative effects of
humidity and of minimum temperature (with a lag of 2 days), which indicates that
higher humidity and higher minimum temperature reduce the expected number of
deaths. Regarding the days of the week, mortality was higher on Monday than on
Sunday and was even lower on all other days. This result might be due to
different behavior and thus personal exposure to the pollutant on weekends or,
more likely, due to a lag in recording on weekends.

\begingroup
\setkeys{Gin}{width=0.45\textwidth}

\begingroup
\begin{figure*}
  \centering

\subfigure{(a)\label{fig:sao_paulo_comp_a}
  \raisebox{-\height}{
\includegraphics{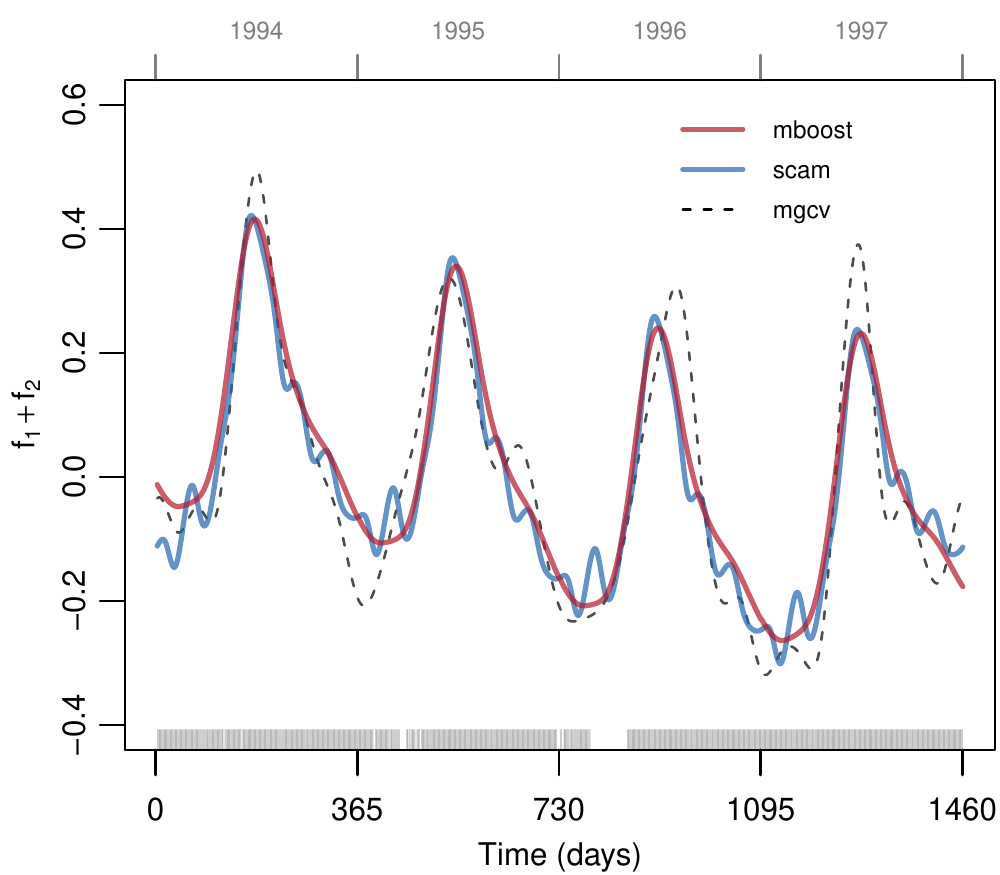}
}}
\hfill
\subfigure{(b)\label{fig:sao_paulo_comp_b}
  \raisebox{-\height}{
\includegraphics{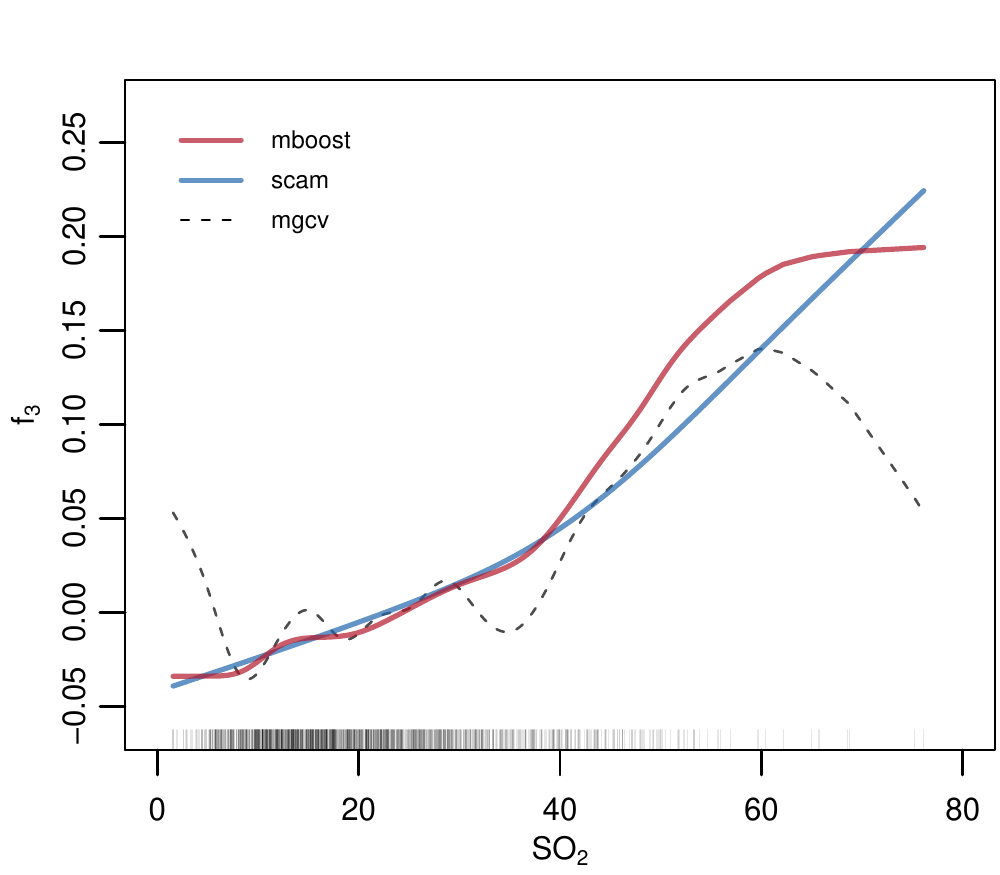}
}}
\caption{Comparison of \subref{fig:sao_paulo_comp_a} the effect estimates for
  time (combination of long-term trend with the seasonal pattern) and
  \subref{fig:sao_paulo_comp_b} the effect estimates for SO$_2$. All effects are
  centered around zero.} \label{fig:sao_paulo_comp}
\end{figure*}

\endgroup

\endgroup

The resulting time trend of \mgcv{} is very similar compared to that of the
model \mbmono{} (Fig.~\ref{fig:sao_paulo_comp_a}), despite the fact that \mgcv{}
was fitted without a cyclic constraint and thus allowing for a changing shape
from year to year. However, the estimation of the complete time pattern without
decomposition into the trend effect and the periodical, seasonal effect was less
stable. The model \scam{} decomposes the seasonal pattern in the same manner as
\mbmono{} and uses a cyclic constraint for the day of the year and a smooth long
term trend. Yet, the resulting effect estimate seems quite unstable at the
boundary of the years where it shows some extra peaks. Modeling the trend and
the periodic effect separately may have the disadvantage that some of the
small-scale changes (e.g., around day 730) are missed. However, without this
decomposition, models do not allow a direct inspection of the seasonal effect
throughout the year. Hence, decomposing influence of time into seasonal effects
and smooth long-term effects seems highly preferable as it offers a stable, yet
flexible method to model the data and allows an easier and more profound
interpretation.

Both monotonic approaches (\mbmono{} and \scam{}) showed a similar pattern in
the effect of SO$_2$ (Fig.~\ref{fig:sao_paulo_comp_b}), but \scam{} does not
flatten out for high values of the SO$_2$ concentration. Estimates from
non-monotonic model (\mgcv{}) were very wiggly for small values up to a
concentration of $40\,\mu$g/m$^3$ and drop to zero for large values, which seems
unreasonable.

Finally, all considered models had almost the same linear effects for the
covariates (results not shown here; see \textsf{R} code in the electronic
supplement). Hence, we can conclude that the linear effects in this model are
very stable and are hardly influenced by the fitting method, i.e., boosting,
penalized iteratively weighted least-squares \citep[P-IWLS; see
e.g.,][]{Wood:2008}, or Newton-Raphson \citep{PyaWood:scam:2014}, nor by
constraints, i.e., monotonic or cyclic constraints, that are used to model the
data.

Concerning the predictive accuracy, we compared the (negative) predictive
log-likelihood of the two constrained models \mbmono{} and \scam{} on 100
bootstrap samples. Both models performed almost identical with an average
predictive risk of $1352.7$ (sd: $34.00$) for \mbmono{} and of $1353.0$ (sd:
$33.74$) for \scam{}. Thus, in this low dimensional example boosting can well
compete with standard approaches for constrained regression. In situations where
variable selection is of major importance or when we want to fit models without
assuming an exponential family, boosting shows its special strengths.

\subsection[Activity of Roe Deer]{Activity of Roe Deer \roedeer{}}
\label{sec:roe-deer-activity}

In the Bavarian Forest National Park (Germany), the applied wildlife management
strategy is regularly examined. Part of the strategy involves trying to
understand the activity profiles of the various species, including lynx, wild
boar, and roe deer. The case study here focuses on the activity of European roe
deer \roedeer{}. According to \citet{Stache:2013}, animal activity is influenced
by exogenous factors, such as the azimuth of the sun (i.e., day/night rhythm and
seasons), temperature, precipitation, and depth of snow. Another important role
is played by endogenous factors, such as the species (e.g., reflected in their
diet; roe deer are browsers), age, and sex. Additionally, as roe deer tend to be
solitary animals, a high level of individual specific variation in activity is
to be expected. The activity data was recorded using telemetry collars with an
acceleration sensor unit. The activity is represented by a number ranging from 0
to 510, where higher values represent higher activity.

Activity profiles for the day and for the year were provided. As earlier
analysis showed, the activity of males and females differs greatly. Hence, sex
should be considered as an effect modifier in the analysis by defining
sex-specific activity profiles. We considered a Gaussian model with the additive
predictor
\begin{equation*}
  \begin{split}
  \mathds{E}(\text{activity}| \cdot) =\; & \x^\top \bv +
         \text{temp} \cdot f_1(t_{\text{days}}) \\
       &+ \text{depth of snow} \cdot f_2(t_{\text{days}})\\
       &+ \text{precipitation} \cdot f_3(t_{\text{days}})\\
       &+ f_4(t_{\text{hours}}, t_{\text{days}})\\
       &+ I_{(\text{sex = male})} f_5(t_{\text{hours}}, t_{\text{days}}) + b_{\text{roe}},
  \end{split}
\end{equation*}
where $\x$ contains the categorical covariates sex, type of collar, year of
observation, and age. Temperature, depth of snow, and precipitation entered the
model rescaled to $|x| \leq 1$ by dividing the variables by the respective
absolute maximum values. The effects of temperature ($f_1$), depth of snow
($f_2$), and precipitation ($f_3$) depend on the calendar day
($t_{\text{days}}$). An interaction surface ($f_4$) for time of the day
($t_{\text{hours}}$) and calendar day ($t$) was specified to flexibly model the
daily activity profiles throughout the year. An additional effect for male roe
deer was specified with $f_5$. Finally, a random intercept $b_{\text{roe}}$ for
each roe deer was included. Details on the data set can be found in
\citetappendix{sec:roedeer_data}, where we also describe the model specification
in detail. \textsf{R} code is given as electronic supplement.

\paragraph{Results}

In the resulting model, six of ten base-learners were selected. The largest
contribution to the model fit was given by the smooth interaction surfaces $f_4$
and $f_5$, which represent the time-dependent activity profiles for male and
female roe deer (Figure~\ref{fig:roedeer_time}). The individual activity of the
roe deer $b_\text{roe}$ substantially contributed to the total predicted
variation, with a range of approximately 20 units (not depicted here). The
time-varying effects of temperature and depth of snow and the effect of the type
of collar had a lower impact on the recorded activity of roe deer.

\begingroup
\setkeys{Gin}{width=0.4\textwidth}
%
%

\begin{figure*}[ht!]
  \centering
  \subfigure{(a)\label{fig:roedeer_time_a}
    \raisebox{2em}{\raisebox{-\height}{
\includegraphics{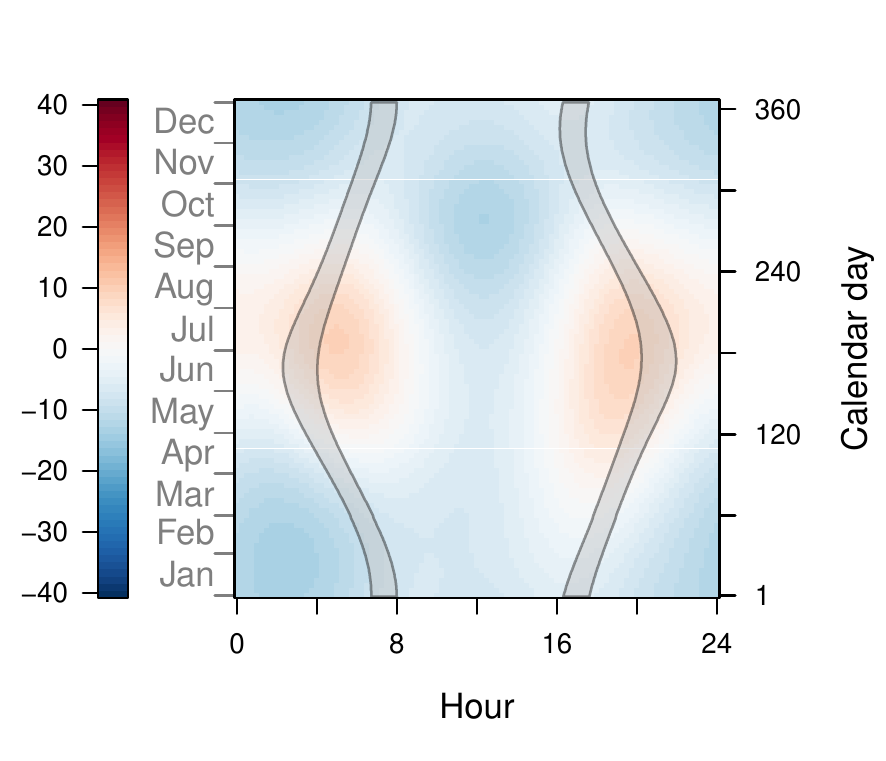}
}}}
\hspace{1cm}
  \subfigure{(b)\label{fig:roedeer_time_b}
    \raisebox{2em}{\raisebox{-\height}{
\includegraphics{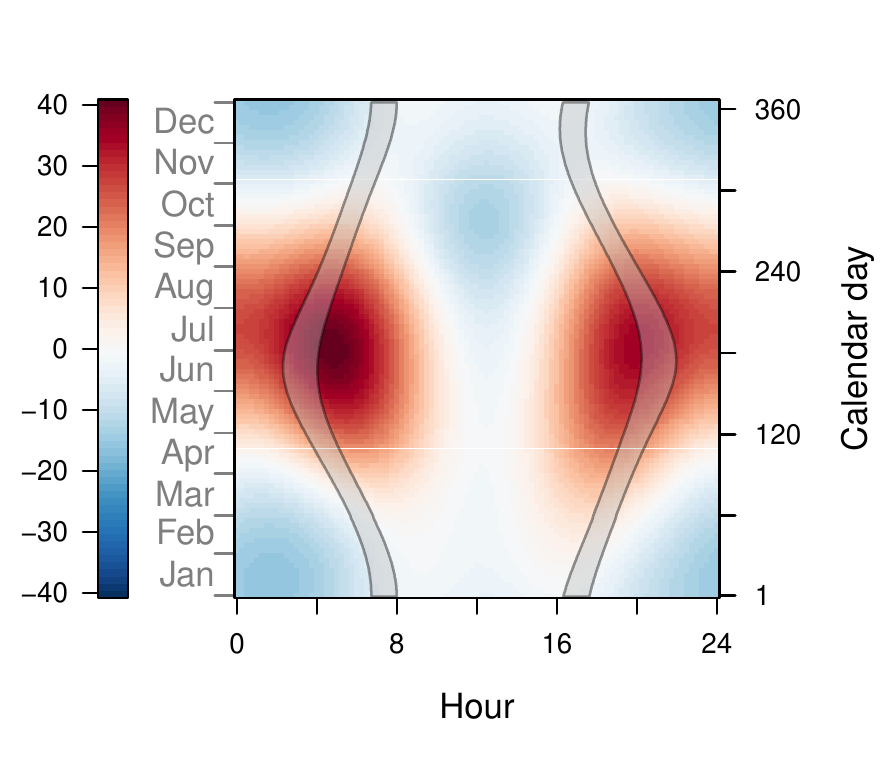}
}}}
\caption{Influence of time on roe deer activity. Combined effect of calendar day
  and time of day \subref{fig:roedeer_time_a} for female roe deer (= $f_4$) and
  \subref{fig:roedeer_time_b} for male roe deer ($= f_4 + f_5$), together with
  twilight phases (gray). White areas depict the mean activity level throughout
  the year; blue shading represents decreased activity, and red shading
  represents increased activity. \label{fig:roedeer_time}}
\end{figure*}

\endgroup

The activity profiles (Fig.~\ref{fig:roedeer_time}) showed that roe deer were
most active in and around the twilight phases in the mornings and evenings. This
holds for the whole year and for both males and females. In general, the
activity profiles of female and male roe deer were very similar, but male
activity was much higher and had more variability. The activity of roe deer was
strongly influenced by the season: During summer, the activity was much higher
throughout the entire day. The phase of least activity was around noon. This
behavior was enhanced in autumn. In spring, activity was more evenly distributed
throughout the daytime, and lowest activity occurred during the hours after
midnight.

The effects of climatic variables are depicted in
Figure~\ref{fig:roedeer_covars}. A higher temperature led to lower activity
(negative effect of temperature), except from May to July, when higher
temperatures led to higher activity (positive effect of temperature). The depth
of snow had a negative effect on roe deer activity throughout the year, i.e.,
deeper snow led to lower activity. The effect of snow depth was stronger in the
summer months (when there is hardly any snow), and less strong in January and
February even though the snow depth was the greatest. Precipitation had no
effect on roe deer activity according to our model.

\begingroup
\setkeys{Gin}{width=0.4\textwidth}

\begin{figure}[ht!]
  \centering
  \subfigure{(a)\label{fig:roedeer_covars_a}
  \raisebox{-\height}{
\includegraphics{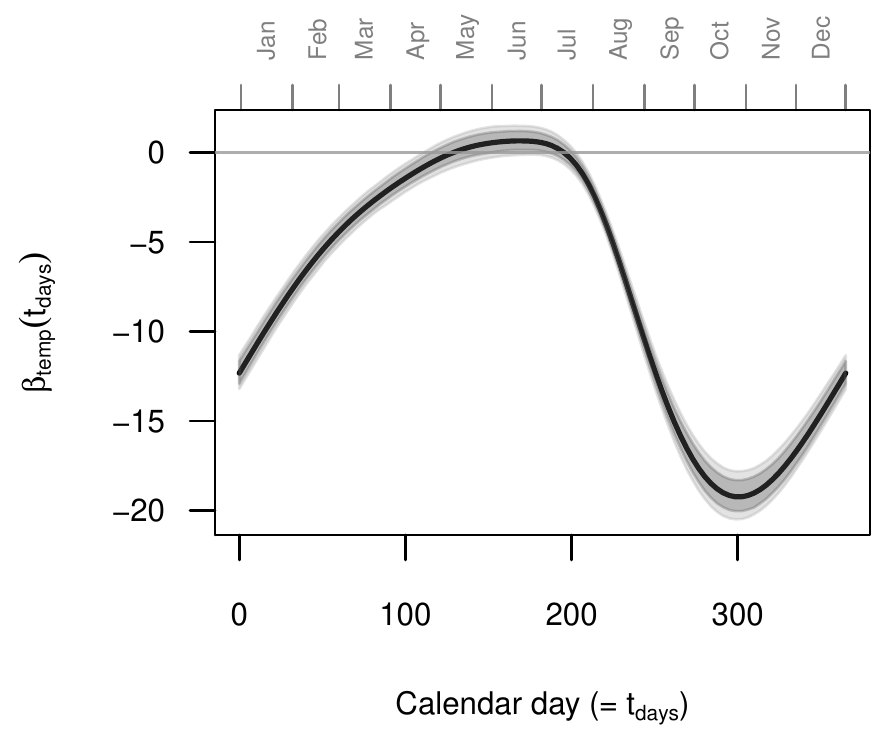}
}}
\hfill
  \subfigure{(b)\label{fig:roedeer_covars_b}
  \raisebox{-\height}{
\includegraphics{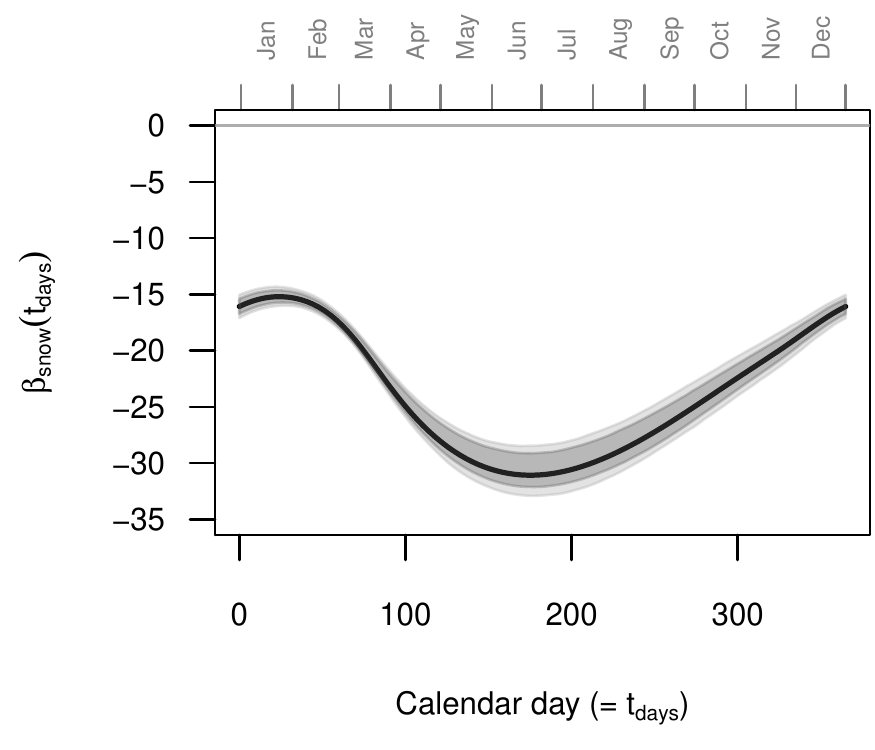}
}}
\caption{Time-varying effects (i.e., $\beta (t_{\text{days}})$) of
  \subref{fig:roedeer_covars_a} temperature and \subref{fig:roedeer_covars_b}
  depth of snow together with 80\% (dark gray) and 95\% (light gray) point-wise
  bootstrap confidence intervals. Note that both variables are rescaled, i.e.,
  $\beta(t_{\text{days}})$ is the maximal effect. On a given day, the effects of
  temperature and depth of snow are linear. The higher the amplitude for a given
  day, the stronger the effect will be. \label{fig:roedeer_covars}}
\end{figure}


\endgroup

\subsection{Deer--vehicle Collisions in Bavaria}
\label{sec:ctm}

Important areas of application for both monotone and cyclic base-learners are
conditional transformation models \citep{Hothorn_Kneib_Buehlmann_2014}. Here, we
describe the distribution of the number of deer--vehicle collisions (DVC) that
took place throughout Bavaria, Germany, for each day $k = 1, \dots, 365$ of the
year 2006 \citep[see][for a more detailed description of the
data]{Hothorn_Brandl_Mueller_2012}, i.e.,
\begin{eqnarray*}
\Prob(\text{number of DVCs} \le y | \text{day} = k) = \Phi(h(y | k)),
\end{eqnarray*}
where $\Phi$ is the distribution function of the standard normal
distribution. The conditional transformation function $h$ is parametrized as
\begin{eqnarray*}
h(y | k) = (\B_\text{day}(k) \otimes \B_\text{DVC}(y)) \bv,
\end{eqnarray*}
where $\B_\text{day}$ is a cyclic B-spline transformation for the day of the
year (where Dec 31 and Jan 1 should match) and $\B_\text{DVC}$ is a B-spline
transformation for the number of deer--vehicle collisions. The Kronecker product
$\B_\text{day}(k) \otimes \B_\text{DVC}(y)$ defines a bivariate tensor product
spline, which is fitted under smoothness constraints in both dimensions. Since
the transformation function $h(y | k)$ must be monotone in $y$ for all days $k$
(otherwise $\Phi(h(y | k))$ is not a distribution function), a monotonicity
constraint is needed for the second term, i.e., one requires monotonicity with
respect to the number of deer--vehicle collisions but not with respect to time.
The model was fitted by minimizing a scoring rule for probabilistic forecasts
(e.g., Brier score or log score). Here, we applied the boosting approach
described in \citep{Hothorn_Kneib_Buehlmann_2014}. \textsf{R} code to fit the
model is given as electronic supplement.

\paragraph{Results} We display the corresponding quantile functions over the
course of the year 2006 in Figure~\ref{fig:DVC}. Three peaks (territorial
movement at beginning of May, rut at end of July/ beginning of August, and early
in October) were identified. While the first two peaks are expected, the
significance of the third peak in October remains to be discussed with
ecologists. Over the year, not only the mean but also higher moments of the
distribution of the number of deer--vehicle collisions varied over time.

\begingroup
\setkeys{Gin}{width=0.45\textwidth}

\begin{figure}
\begin{center}
\includegraphics{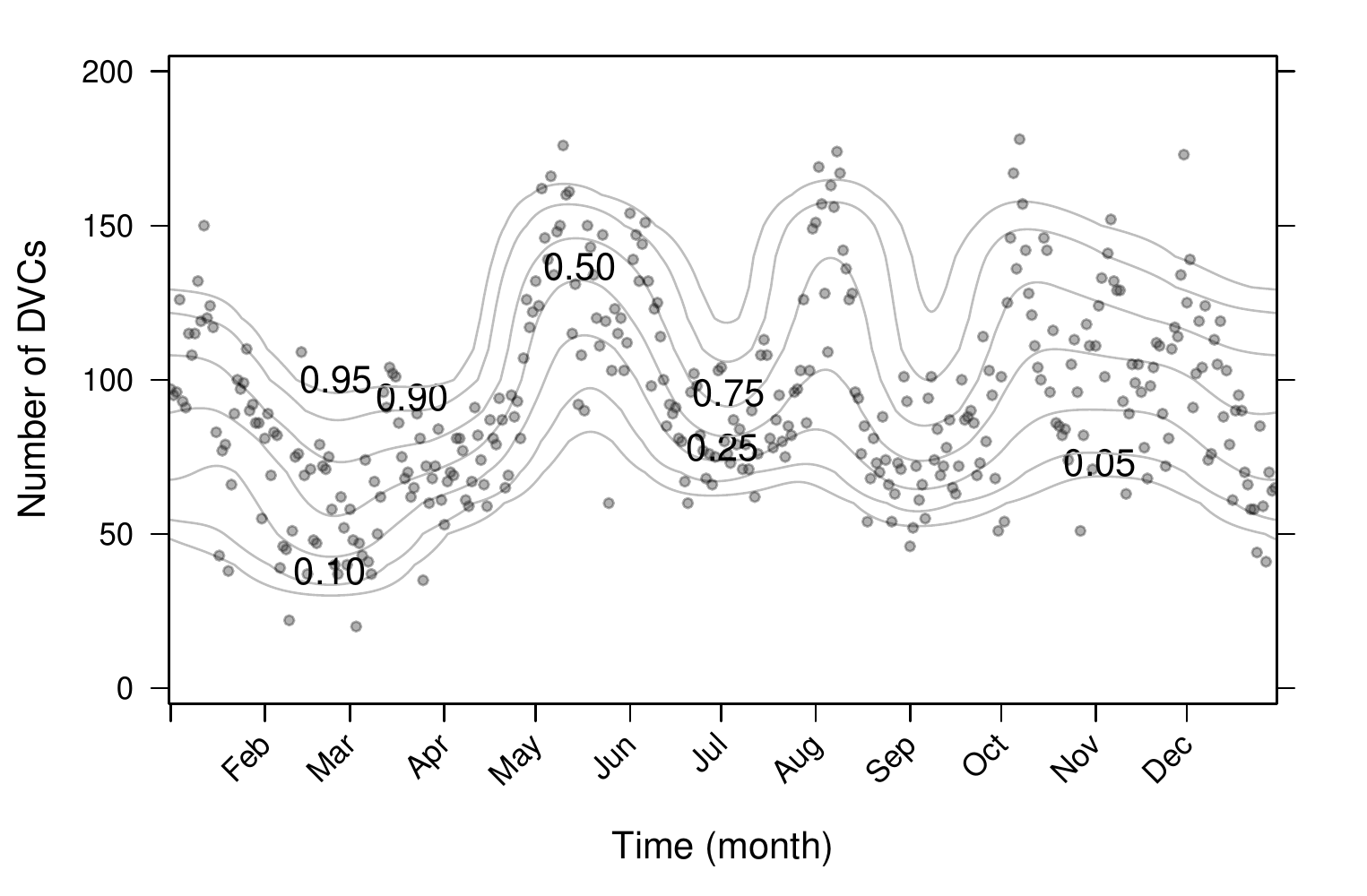}
\caption{Number of deer--vehicle collisions (DVCs) in Bavaria, Germany, for each
  day of the year 2006. Superimposed lines depict the conditional quantiles
  ($5\%, 10\%, 25\%, 50\%, 75\%, 90\%, 95\%$) of this distribution for each day
  of the year. \label{fig:DVC}}
\end{center}
\end{figure}

\endgroup

\section{Concluding Remarks}

In this article, we extended the flexible modeling framework based on boosting
to allow inclusion of monotonic or cyclic constraints for certain variables.

The monotonicity constraint on continuous variables leads to monotonic, yet
smooth effects. Monotonic effects can be furthermore applied to bivariate
P-splines. In this case, one can specify different monotonicity constraints for
each variable separately. However, it is ensured that the resulting interaction
surface is monotonic (as specified). Monotonicity constraints might be
especially useful in, but are not necessarily restricted to, data sets with
relatively few observations or noisy data. The introduction of monotonicity
constraints can help to estimate more appropriate models that can be
interpreted. In the context of conditional transformation models monotonicity
constraints are an essential ingredient as we try to estimate distribution
functions, which are per definition monotonic. Many other approaches to
monotonic modeling result in non-smooth function estimates
\citep[e.g.,][]{Detteetal:Mono:2006,Leeuw_Pava:2009,FangMeinshausen:LISO:2012}.
In the context of many applications, however, we feel that smooth effect
estimates are more plausible and hence preferable.

Cyclic estimates can be easily used to model, for example, seasonal effects. The
resulting estimate is a smooth effect estimate, where the boundaries are
smoothly matched. Cyclic effects can be applied straightforwardly to model
surfaces where the boundaries in each direction should match if cyclic tensor
product P-splines are used. The idea of cyclic effects could also be extended to
ordinal covariates with a temporal, periodic effect --- such as days of the
week.

Finally, both restrictions --- monotonic and cyclic constraints --- can be mixed
in one model: Some of the covariates are monotonicity restricted, others have
cyclic constraints and the rest is modeled, for example, as smooth effects
without further restrictions or as linear effects.

Both monotonic P-splines and cyclic P-splines integrate seamlessly in the
functional gradient descent boosting approach as implemented in \textbf{mboost}
\citep{Hofner:mboost:2014,pkg:mboost:CRAN:2.4}. This allows a single framework
for fitting possible complex models. Additionally, the idea of asymmetric
penalties for adjacent coefficients can be transferred from P-splines to ordinal
factors \citep{Hofner:monotonic:2011}, which can be integrated in the boosting
framework as well. (Constrained) boosting approaches can be used in any
situation where standard estimation techniques are used. They are especially
useful, when variable and model selection are of major interest, and they can be
used even if the number of variables is much larger than the number of
observations ($p \gg n$). The proposed framework can be used to fit generalized
additive models if one uses the negative log-likelihood as loss function. Other
loss functions to fit constrained quantile or expectile regression models
\citep{Fenske:2011,Sobotka:2012} or robust models with constraints based on the
Huber loss can be used straight forward. The framework can be also transferred
to conditional transformation models \citep{Hothorn_Kneib_Buehlmann_2014} or
generalized additive models for location, scale and shape
\citep[GAMLSS;][]{Rigby:gamlss:2005}, where boosting methods where recently
developed \citep{mayr:gamboostlss:2012,Hofner:gamboostLSS_tutorial:2014}.

\begin{acknowledgements}
  We thank the ``Laborat\'{o}rio de Polui\c{c}\~{a}o Atmos\-f\'{e}rica
  Experimental, Faculdade de Medicina, Universidade de S\~{a}o Paulo, Brasil'',
  and Julio M. Singer for letting us use the data on air pollution in S\~{a}o
  Paulo. We thank Marco Heurich from the Bavarian Forest National Park,
  Grafenau, Germany, for the roe deer activity data, and Karen A. Brune for
  linguistic revision of the manuscript. We also thank the Associate Editor and
  two anonymous reviewers for their stimulating and helpful comments.
\end{acknowledgements}

\bibliographystyle{spbasic}
\bibliography{bibliography}

\end{document}